\documentclass[letterpaper, 10 pt, conference]{ieeeconf}  

\IEEEoverridecommandlockouts                              

\usepackage[utf8]{inputenc}
\usepackage{amsmath,amssymb}
\usepackage{fancyvrb}
\usepackage{times}
\usepackage{wrapfig}
\usepackage{graphicx}
\usepackage{xcolor}
\usepackage{xspace}
\usepackage{array}
\usepackage{tabulary}
\usepackage{multirow}
\usepackage{color, colortbl}
\definecolor{Gray}{gray}{0.2}
\definecolor{LightRed}{rgb}{1.0, 0.5, 0.5}
\definecolor{LightYellow}{rgb}{1.0, 1.0, 0.13}
\definecolor{LightOrange}{rgb}{1.0, 0.46, 0.09}
\definecolor{LighterYellow}{rgb}{1.0, 1.0, 0.5}
\definecolor{LighterOrange}{rgb}{1.0, 0.7, 0.3}

\makeatletter
\let\NAT@parse\undefined
\makeatother
\usepackage{hyperref}

\graphicspath{{figures/}}

\newcommand{\scenic}{\textsc{Scenic}\xspace}
\newcommand{\verifai}{\textsc{VerifAI}\xspace}

\newcommand{\Scenario}{\ensuremath{\mathcal{S}}}
\newcommand{\testcase}{x}
\newcommand{\testrun}{r}
\newcommand{\tr}{\ensuremath{\tau}}
\newcommand{\simModel}{\ensuremath{\mathcal{M}}}
\newcommand{\params}{\ensuremath{\vec{\pi}}}
\newcommand{\pgm}{P}

\newcommand{\startdistance}{\ensuremath{d_{\text{start}}}}
\newcommand{\startdelay}{\ensuremath{t_{\text{start}}}}
\newcommand{\walkdistance}{\ensuremath{d_{\text{walk}}}}
\newcommand{\hesitatetime}{\ensuremath{t_{\text{hesitate}}}}

\newcommand{\collisionspec}{\ensuremath{\varphi_{\text{safe}}}}

\DefineVerbatimEnvironment{bscenario}{BVerbatim}{samepage=true,fontfamily=cmtt}
\DefineVerbatimEnvironment{scenario}{Verbatim}{numbers=left,xleftmargin=7mm,numbersep=4mm,samepage=true,fontfamily=cmtt}

\newcounter{myctr}
\newenvironment{mylist}{\begin{list}{\arabic{myctr}.}
{\usecounter{myctr}
\setlength{\topsep}{1mm}\setlength{\itemsep}{0.25mm}
\setlength{\parsep}{0.1mm}
\setlength{\itemindent}{0mm}\setlength{\partopsep}{0mm}
\setlength{\labelwidth}{15mm}
\setlength{\leftmargin}{4mm}}}{\end{list}}

\newenvironment{myitemize}{\begin{list}{$\bullet$}
{\setlength{\topsep}{1mm}\setlength{\itemsep}{0.25mm}
\setlength{\parsep}{0.1mm}
\setlength{\itemindent}{0mm}\setlength{\partopsep}{0mm}
\setlength{\labelwidth}{15mm}
\setlength{\leftmargin}{4mm}}}{\end{list}}

\renewcommand{\baselinestretch}{0.948}

\title{\LARGE \bf
Formal Scenario-Based Testing of Autonomous Vehicles:\\ From Simulation to the Real World
}

\author{Daniel J. Fremont$^{1,4}$, 
Edward Kim$^{1}$,
Yash Vardhan Pant$^{1}$,
Sanjit A. Seshia$^{1}$,\\
Atul Acharya$^{2}$, 
Xantha Bruso$^{2}$, 
Paul Wells$^{2}$, 
Steve Lemke$^{3}$, Qiang Lu$^{3}$,
Shalin Mehta$^{3}$
\thanks{\noindent$^{1}$University of California, Berkeley}
\thanks{$^{2}$American Automobile Association (AAA) of Northern California, Nevada \& Utah (NCNU)}%
\thanks{$^{3}$LG Electronics America R\&D Lab}%
\thanks{$^{4}$University of California, Santa Cruz}%
\thanks{This work has been supported in part by the National Science Foundation Graduate Research Fellowship Program under Grant No. DGE-1752814, NSF grants CNS-1545126 (VeHICaL), CNS-1646208, and CNS-1837132, the DARPA Assured Autonomy program, the Collaborative Sciences Center for Road Safety (CSCRS) under Grant No. 69A3551747113, Berkeley Deep Drive, and Toyota through the iCyPhy center.}
}

\begin{document}

\maketitle
\thispagestyle{empty}
\pagestyle{empty}

\begin{abstract}
We present a new approach to automated scenario-based testing of the safety of autonomous vehicles, especially those using advanced artificial intelligence-based components, spanning both simulation-based evaluation as well as testing in the real world. 
Our approach is based on formal methods, combining 
formal specification of scenarios and safety properties,
algorithmic test case generation using formal simulation, 
test case selection for track testing,
executing test cases on the track, and 
analyzing the resulting data.
Experiments with a real autonomous vehicle at an industrial testing facility 
support our hypotheses that 
(i) formal simulation can be effective at identifying test cases to run
on the track, and
(ii) the gap between simulated and real worlds can be systematically evaluated
and bridged.
\end{abstract}

\section{Introduction}\label{sec:intro}

A defining characteristic of the growth in autonomous vehicles (AVs) and automated driving systems (ADS)
is the expanding use of machine learning (ML) and other artificial intelligence (AI) based components in them.
ML components, such as deep neural networks (DNNs), have proved to be fairly effective at perceptual tasks,
such as object detection, classification, and image segmentation, as well as for prediction of agent behaviors.
However, it is known that ML components can be easily fooled by so-called adversarial examples,
and there have also been well-documented failures of AVs in the real world for which the evidence points to
a failure (in part) of ML-based perception.
Therefore, there is a pressing need for better techniques for testing and verification of ML/AI-based ADS and AVs~\cite{seshia-arxiv16}.

Simulation is regarded as an important tool in the design and testing of AVs with ML components. 
Advanced photorealistic simulators are now available for AVs, providing designers with the ability to simulate
``billions of miles'' so as to test their AV components, cover corner-case scenarios that are hard to test
in the real world, and diagnose issues that arise in real-world testing, such as disengagements. 
However, some key questions remain.
How well does simulation match the real world? What is the
value of simulation vis-a-vis testing in a physical environment that includes other vehicles, pedestrians,
and other road users?

\begin{figure}[t]
    \centering
	\includegraphics[width=\columnwidth]{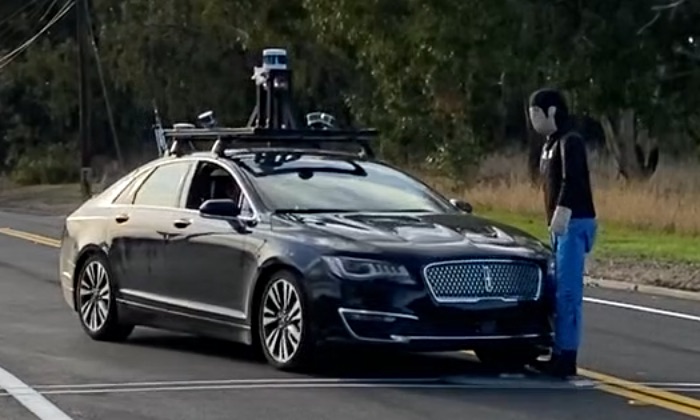}
	\caption{The Autonomous Vehicle (AV) and pedestrian dummy used for track testing. The picture shows the AV hitting the pedestrian during testing (test F1 Run 1, \url{https://youtu.be/PehgLCGHF5U}), see Section \ref{sec:track} for details.}
	\label{fig:LG_Car}
\end{figure}

An intermediate step between simulation and testing on public roads is {\em track testing}. This form of testing
involves driving the AV on roads in a test facility with a reasonable degree of control over the other agents around the AV,
including, for example, pedestrian dummies and inflatable cars to use for crash testing.
Track testing allows one to run the actual AV with its real hardware and software systems in environments
that can be designed to mimic certain challenging driving conditions.
However, track testing can be very expensive, labor-intensive, and time-consuming to set up.
Given these challenges, which tests should one run?
For testing AVs with complex ML-based components, it is crucial to be able to run the tests that
will prove most effective at identifying failures or strange behavior, uncovering bugs, and increasing
assurance in the safety of the AV and ADS.

This paper takes a step towards addressing these problems by investigating the following two questions:
\begin{mylist}
    \item 
  {\it Can formal simulation aid in designing effective road tests for AVs?}
  By {\em formal simulation} we mean simulation-based testing that is guided by the use of
  formal models of test scenarios and formal specification of safety properties and metrics.
  More specifically, do unsafe (safe) runs in simulation produce unsafe (safe) runs on the track?
  How should one select tests from simulation to run on the track?

\item 
{\it How well can simulation match track testing of AVs?}
We aim to quantitatively and qualitatively compare simulation and track testing data,
for a test scenario that has been formally specified and implemented in both
simulation and track testing.
\end{mylist}

Our approach is rooted in formal methods, a field centered on the use of mathematical models of systems and their
requirements backed by computational techniques for their design and verification.
In particular, we use a formal probabilistic programming language, \scenic~\cite{scenic}, to specify a test scenario, encapsulating
key behaviors and parameters of the AV and its environment.
Additionally, we use formal specification languages, such as Metric Temporal Logic~\cite{mtl}, to specify
safety properties for AVs.
We combine formally-specified scenario descriptions and requirements with algorithms for 
simulation-based verification of AVs, also known as {\em falsification}, implemented in an open-source
toolkit called \verifai~\cite{verifai}.
These methods, when combined with an advanced photorealistic full-stack simulator for AVs, the LGSVL Simulator~\cite{lgsvl}, 
allow us to identify safe and unsafe behaviors of the AV in simulation.
We seek to answer the above questions by incorporating into the simulator a ``digital twin'' of an industrial-scale
test track, the GoMentum Station test facility in Concord, California~\cite{gomentum-www}.
We have developed and deployed a simulation-to-test-track flow where formal simulation is used to identify test
cases to execute on the track, and these test cases are systematically mapped onto hardware that is used to control
agents on the track in the AV's environment.
We present the results of executing this flow on a scenario involving a pedestrian crossing in front of an AV,
providing evidence for the effectiveness of formal simulation for identifying track tests,
as well as a quantitative mechanism for comparing simulation results with those obtained on the track.
Specifically, our results indicate that:
\begin{mylist}
\item Our formal simulation-based approach is effective at synthesizing test cases that transfer well to the track: 62.5\% of unsafe simulated test cases resulted in unsafe behavior on the track, including a collision; 93.3\% of safe simulated test cases resulted in safe behavior on the track (and no collisions).
Our results also shed light on potential causes for AV failure in perception, prediction, and planning.

\item While AV and pedestrian trajectories obtained in simulation and real-world testing for the same test were qualitatively similar
(e.g., see Fig~\ref{fig:trajectories}), we also noted differences as quantified using time-series metrics~\cite{dtw,skorokhod} and with metrics
such as minimum distance between the AV and pedestrian. Variations exist even amongst simulations of the same test case due to non-deterministic behavior of the AV stack, although these are smaller.

\end{mylist}

\subsection*{Related Work}

Scenario-based testing of AVs is a well-studied area. One approach is to construct
tests from scenarios created from crash data analysis (\cite{crash_data1,crash_data2}) and naturalistic driving data (NDD) analysis (\cite{ndd_2,ndd_1}), which leverage human driving data to generate test scenarios. Similarly, the PEGASUS project~\cite{Pegasus2019} focuses on (i) bench-marking human driving performance using a comprehensive dataset comprising crash reports, NDD, etc., and (ii) characterizing the requirements that AVs should satisfy to ensure the traffic quality is at least unaffected by their presence. 
Our work differs from these in the use of formal methods for specifying scenarios and safety properties,
as well as in automated synthesis of test cases.

Our use of a scenario specification language, \scenic~\cite{scenic}, is related to other work on scenario description languages. 
OpenSCENARIO~\cite{openscenario} defines a file format for the description of the dynamic content of driving and traffic simulators, based on the extensible markup language (XML).
GeoScenario~\cite{geoscenario} is a somewhat higher-level domain specific language (DSL) for scenario representation, whose syntax also looks like XML.
\scenic is a flexible high-level language that is complementary to these.
The Measurable Scenario Description Language (M-SDL)~\cite{msdl} is a recent higher-level DSL
similar to \scenic, which precedes its definition; while M-SDL is more specialized for AV testing,
it has less support than \scenic for probabilistic and geometric modeling and is not supported by
open-source back-end tools for verification, debugging, and synthesis of autonomous AI/ML based systems,
unlike \scenic which is complemented by the open-source \verifai toolkit~\cite{verifai}.

Recent work on the test scenario library generation (TSLG) problem (\cite{Feng1,Feng2}) mathematically describes a scenario, defines a relevant metric, and generates a test scenario library. A critical step in TSLG is to construct a surrogate model of an autonomous vehicle. The authors construct this based on human driving data, which, while useful, may not capture the subtleties in complex ML/AI-based autonomous vehicle stacks. Additionally, the work presents only simulation results, whereas our paper reports on both simulation and track testing with a real AV.
Abbas et al.~\cite{abbas2017safe} present a test harness for testing an AV's perception and control stack in a simulated environment and searching for unsafe scenarios. However, in the absence of a formal scenario description language, representing an operational design domain (ODD) becomes tedious manual labor and challenging as the number of traffic participants scales up. 

Researchers have considered the gap between simulation and road/track testing. 
A methodology for testing AVs in a closed track, as well as in simulation and mixed-reality settings, is explored in \cite{AntkiewiczSAE2020}. The main aim there is to evaluate the AV's performance across the different settings using standard tests \cite{WISE2018}, rather than use computational techniques to generate tests based on formally-specified scenarios and outcomes, as we aim to do. 
A recent SAE EDGE research report~\cite{sae-edge} dives deeper into unsettled issues in determining appropriate modeling fidelity for automated driving systems. While it raises important questions, it does not address formal methods for evaluation as we do.

In summary, to the best of our knowledge, this paper is the first to apply a formal methods-based approach to evaluating the safety of ML-based autonomous vehicles
spanning formal specification of scenarios and safety properties, formal simulation-based test generation and selection for track testing, as well as evaluation of the methodology in both simulation and the real world, including systematically measuring the gap between simulation and track testing.

\section{Background}\label{sec:background}

\subsection{\scenic: A Scenario Specification Language}

\scenic{}~\cite{scenic} is a domain-specific probabilistic programming language for modeling the environments of cyber-physical systems.
A \scenic{} program defines a distribution over \emph{scenes}, configurations of objects and agents; a program describing ``bumper-to-bumper traffic'' might specify a particular distribution for the distance between cars, while letting the location of the scene be uniformly random over all 3-lane roads in a city.
\scenic{} provides convenient syntax for geometry, along with declarative constraints, which together make it possible to define such complex scenarios in a concise, readable way.
\scenic{} has a variety of applications to the design of ML-based systems: for example, one can write a \scenic{} program describing a rare traffic scenario like a disabled car blocking the road, then sample from it to generate specialized training data to augment an existing dataset~\cite{scenic}.
More generally, the formal semantics of the language allow it to be used as a precisely-defined model of the environment of a system, as we will see in Sec.~\ref{sec:sim:scenic}.

\subsection{The \verifai Toolkit}

The \verifai{} toolkit~\cite{verifai} provides a unified framework for the design and analysis of AI- and ML-based cyber-physical systems, based on a simple paradigm: simulations driven by formal models and specifications.
In \verifai{}, we first parametrize the space of environments and system configurations of interest, either by explicitly defining parameter ranges or using the \scenic{} language described above.
\verifai{} then generates concrete tests by searching this space, using a variety of algorithms ranging from simple random sampling to global optimization techniques. 
Each test results in a simulation run, where the satisfaction or violation of a system-level specification is checked; the results of each test are used to guide further search, and any violations are recorded in a table for further analysis.
This architecture enables a wide range of use cases, including falsification, fuzz testing, debugging, data augmentation, and parameter synthesis, demonstrated in~\cite{verifai,taxinet}.

\subsection{The LGSVL Simulator}

The LGSVL Simulator~\cite{lgsvl} is an open-source autonomous driving simulator used to facilitate the development and testing of autonomous driving software systems. With support for the Robot Operating System (ROS, ROS2), and alternatives such as CyberRT, the simulator can be used with popular open source autonomous platforms like Apollo (from Baidu) and Autoware (from the Autoware Foundation). 
Thus, it allows one to simulate an entire autonomous vehicle with a full sensor suite in a safe, deterministic (assuming determinism in the AV software stack behavior), and realistic 3D environment. The LGSVL Simulator provides simultaneous real-time outputs from multiple sensors including cameras, GPU-accelerated LiDAR, RADAR, GPS, and IMU. Environmental parameters can be changed including map, weather, time of day, traffic and pedestrians, and the entire simulation can be controlled through a Python API. The LGSVL Simulator is available free and open-source on GitHub (\url{https://github.com/lgsvl/simulator}).

\subsection{GoMentum Station Testing Facility}

GoMentum Station (GoMentum)~\cite{gomentum-www} is currently the largest secure AV test site in the United States. Located in Concord, CA, 35 miles from San Francisco and 60 miles from Silicon Valley, GoMentum features 19 miles of roadways, 48 intersections, and 8 distinct testing zones over 2,100 acres, with a variety of natural and constructed features.
Vehicle manufacturers, AV system developers and other entities have been testing connected and automated vehicles at GoMentum since 2014. 
Our experiments were conducted in the ``urban'' or ``downtown'' zone of GoMentum, which features mostly flat surface streets that are approximately 20 feet wide and several signed, unsigned, and signalized intersections amid an urban and suburban landscape with buildings, trees, and other natural and human-made features. Speeds are restricted to under 30 mph. The roads feature clear and visible lane markings on freshly paved roads. The traffic signs include one-way, stop, yield, and speed limits.

\section{Methodology}\label{sec:method}

We now describe the methodology we use to assess the safety of the AV in simulation,
identify test cases to run at the testing facility (track),  
implement those tests on the real AV and associated testing hardware,
and perform post-testing data analysis.

\begin{figure*}[ht]
    \centering
    \includegraphics[width=0.75\textwidth]{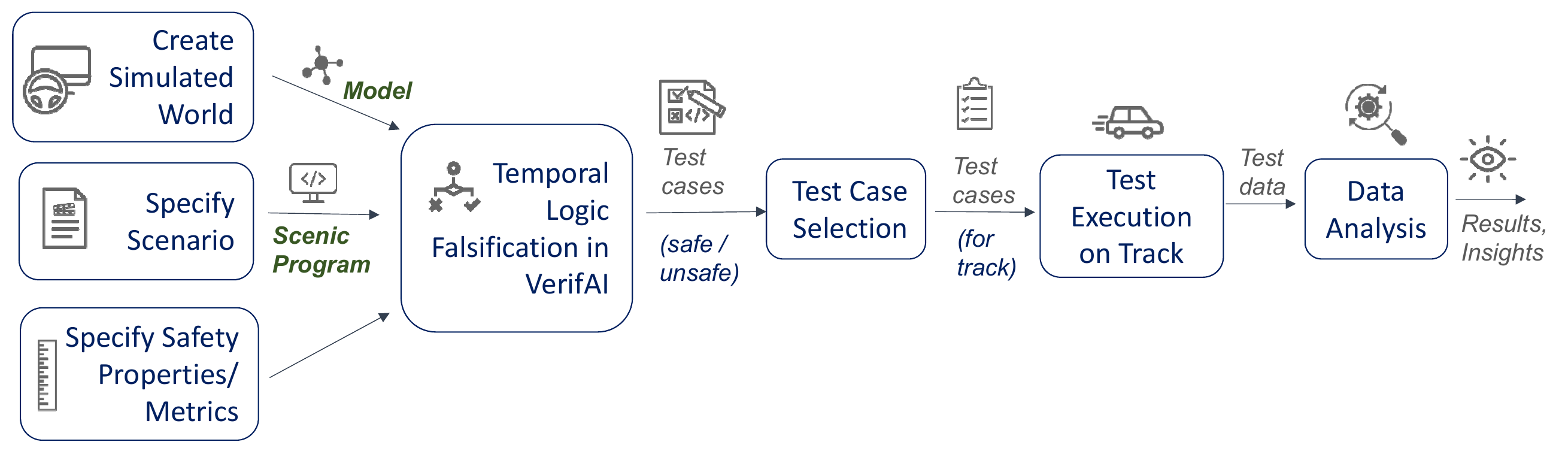}
    \caption{Formal scenario-based testing methodology used in this paper.}
    \label{fig:method}
    \vspace*{-4mm}
\end{figure*}

Let $\simModel$ be the {\it simulation model}, including the full software stack and vehicle
dynamics model of the AV, and its environment, including models of all other objects and agents
in the simulated world.
This model can be configured through a vector of {\it parameters} $\params$,
typically supplied through a configuration file or suitable API to the simulator.
Each valuation of $\params$ defines a {\it test case} $\testcase$.
We assume, for this section, that each test case $\testcase$ produces a unique
{\it simulation run}.\footnote{Note, however, that some industrial simulators and AV stacks 
tend to be non-deterministic
in that the configurable parameters may not define a unique simulation run. We will discuss
later the impact of such non-determinism on our results.}
The time series data generated by the simulation run is referred to as
a {\it trace} $\tr$.
Each test case $\testcase$ is designed so as to also be implementable on the real AV
on the track, although such implementation can be non-trivial as we describe later.
On the track, the environment is less controllable than in the simulator, and therefore
a single test case $\testcase_i$ can produce multiple test runs $\testrun_{i,1}, \testrun_{i,2}, \ldots$.

A key aspect of our method is to formally specify a set of test cases along with an
associated probability distribution over them. We refer to this distribution of test cases
as a {\em scenario} $\Scenario$, which is defined by a \scenic program $\pgm_{\Scenario}$.
Typically, a subset of simulation parameters $\params$ are modeled in $\pgm_{\Scenario}$,
while the others are left fixed for the experiment.

Our overall methodology is depicted in Fig.~\ref{fig:method}, and involves the following steps:
\begin{mylist}
    \item {\em Create Simulation Model} (Sec.~\ref{sec:sim:model}):
    The first step is to create a photorealistic simulation environment including dynamical models for a range of agents implementable on the test track. This involves high-definition (HD) mapping,
    collecting sensor data, using the collected data to create a detailed
    3D mesh, loading that mesh into the simulator, annotating details of drivable areas in the simulator, and combining the resulting 3D world model in the simulator with vehicle and agent dynamics models.
    
    \item {\em Formalize Test Scenario} (Sec.~\ref{sec:sim:scenic}):
    The next step is to formalize the test scenario(s) to execute on the track. We take a formal methods approach, specifying test scenarios in the \scenic probabilistic programming language~\cite{scenic}.
    
    \item {\em Formalize Safety Property/Metric} (Sec.~\ref{sec:sim:scenic}):
    Along with formally specifying the scenario, we must also specify one or more properties that capture the conditions under which the AV is deemed to be operating safely. In formal methods, safety properties over traces are usually specified in a logical notation such as temporal logics. When these properties are quantitative, we term them safety metrics.
    
    \item {\em Identify Safe/Unsafe Test Cases} (Sec.~\ref{sec:sim:verifai}):
    Once the above three steps are complete, the simulation model, test scenario, and safety properties are fed into the \verifai tool to perform falsification. 
    The \scenic{} scenario $\pgm_{\Scenario}$ defines a distribution over the parameters $\params$.
    We configured \verifai{} to sample from this distribution, simulating each corresponding test case and monitoring the safety properties on the resulting trace.
    \verifai{} stores the sampled values of $\params$ in \emph{safe} or \emph{error} tables depending on whether the test satisfies or violates the specification.
    Moreover, \verifai{} uses the \emph{robust semantics} of metric temporal logic (MTL)~\cite{mtl-robustness} to compute a quantitative \emph{satisfaction value} $\rho$ for the specification $\varphi$ which indicates \emph{how strongly} it is satisfied: $\rho > 0$ implies $\varphi$ is satisfied, and larger values of $\rho$ mean that larger modifications to the trace would be necessary for $\varphi$ to instead be falsified.
    The resulting test cases are fed to the next step.

    \item {\em Select Test Cases for Track Testing} (Sec.~\ref{sec:sim:select}):
    \verifai{} provides several techniques, such as Principal Component Analysis and clustering, to automatically analyze the safe and error tables and extract patterns. For low-dimensional spaces, direct visualization can also be used to identify clusters of safe/unsafe tests. Using either approach, we identify different behavior modes, and select representative test cases to execute on the track.
    
    \item {\em Implement Selected Test Cases on Track} (Sec.~\ref{sec:track}):
    Once test cases have been identified in simulation, we need to execute them on the track. For this, dynamic agents (environment vehicles, pedestrians, bicyclists, etc.) must be controllable using parameters (e.g., starting location, time to start motions, velocities, etc.) specified in the \scenic program and synthesized into a test case. Even state-of-the-art hardware for track testing can have limitations that must be matched to the tests synthesized in simulation so as to accurately reproduce them on the track.
    
    \item {\em Record Results and Perform Data Analysis} (Sec.~\ref{sec:analysis}):
    Finally, during track testing, we record various data including videos of the AV moving through the test environment, data on the AV including all sensor data and log data from the AV software stack, as well as data from the test track hardware including GPS beacons and the hardware used to control dynamic agents such as a pedestrian dummy. 
    We then analyze this data to evaluate the effectiveness of test case selection through formal simulation, the correspondence between simulation traces and the traces from track experiments, and potential reasons for unsafe or interesting behavior of the AV.

\end{mylist}

\section{Simulation}\label{sec:sim}
\subsection{Simulation Model Creation}
\label{sec:sim:model}

The photorealistic simulation environment is a ``digital twin'' of the ``Urban A'' test area at GoMentum. The environment was created by collecting hundreds of gigabytes of LiDAR point cloud, camera image, and location data while driving around the site. The collected point cloud data was processed and converted into a unified 3D mesh representing every bump and crack in the road surface as well as all of the surrounding objects including curbs, sidewalks, buildings, signs, etc. Tens of thousands of captured images were then processed into textures and applied to the 3D mesh. The mesh was loaded into the LGSVL Simulator, which was used to annotate details of the drivable areas including lane lines, driving directions, road speeds, crosswalks, intersections, and traffic signs. The annotated, textured mesh was then compiled into a loadable simulation environment along with HD maps that were used both in simulation and in the AV for real-world testing.\footnote{The GoMentum digital twin environment is available as a free download including HD maps in the OpenDRIVE, Lanelet2, Apollo, and Autoware formats at \url{https://content.lgsvlsimulator.com/maps/gomentum/}.}

\subsection{Test Scenario and Safety Properties}
\label{sec:sim:scenic}

\begin{figure}
    \centering
    \includegraphics[width=\columnwidth, trim={2.5cm 0 2.5cm 0},clip]{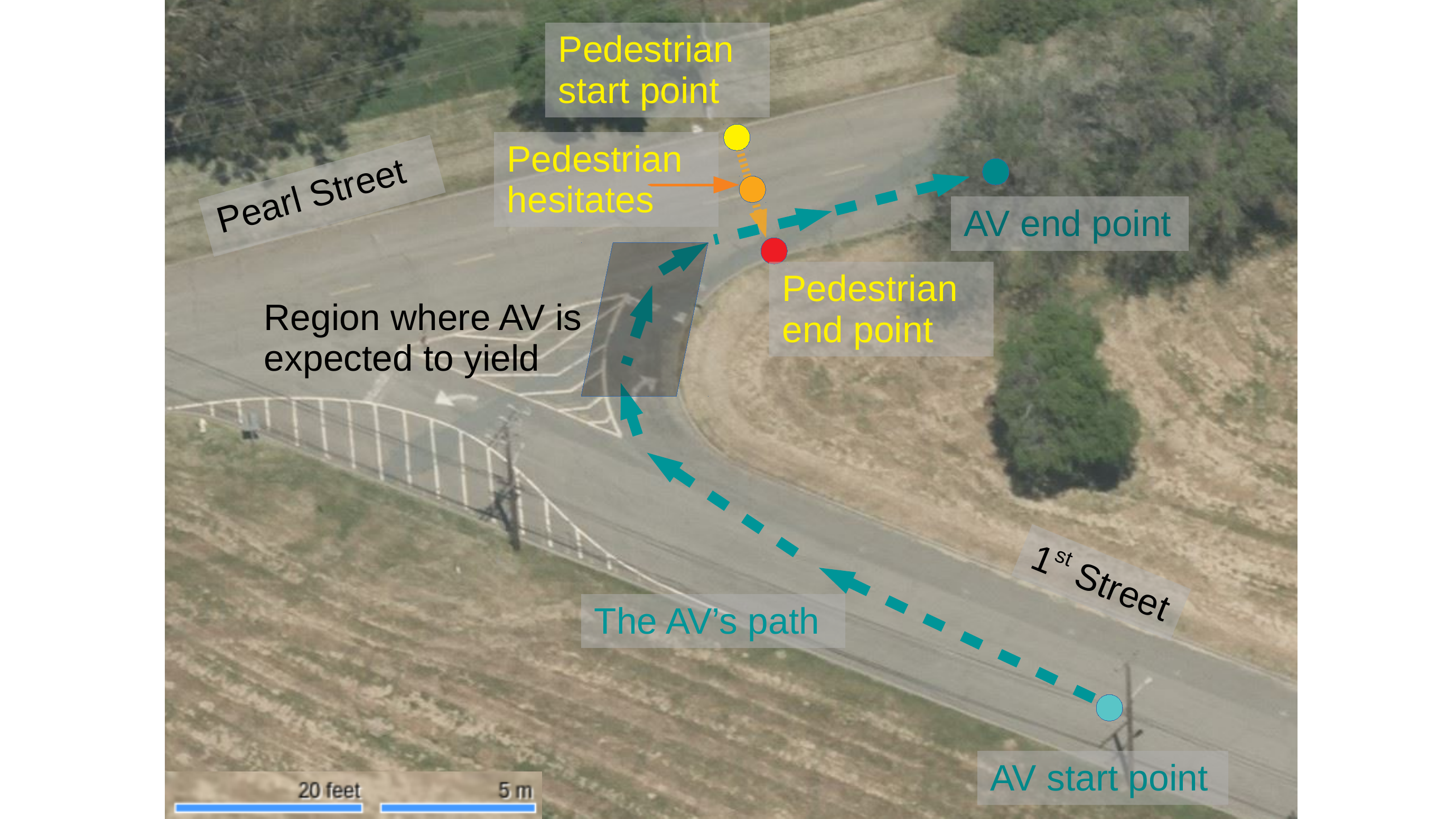}
    \caption{Bird's-eye view of the scenario $\Scenario$.}
    \label{fig:scenario-diagram}
\end{figure}

We selected a scenario where the AV turns right at an intersection, encountering a pedestrian who crosses the road with hesitation, which the AV drives through.
This scenario is diagrammed in Fig.~\ref{fig:scenario-diagram}.
To aid implementation on the track, we fixed the initial positions and orientations of the AV and pedestrian, and defined the pedestrian's trajectory as a straight line with 3 parameters\footnote{Other parametrizations are possible. Our choice here corresponds most directly to what we could implement on the test track: see Sec.~\ref{sec:track-setup}.}:
\begin{itemize}
    \item the delay $\startdelay$ after which the pedestrian starts crossing (with a fixed speed of 1 m/s);
    \item the distance $\walkdistance$ the pedestrian walks before hesitating;
    \item the amount of time $\hesitatetime$ the pedestrian hesitates.
\end{itemize}

\begin{figure}
    \centering
    \includegraphics[width=\columnwidth]{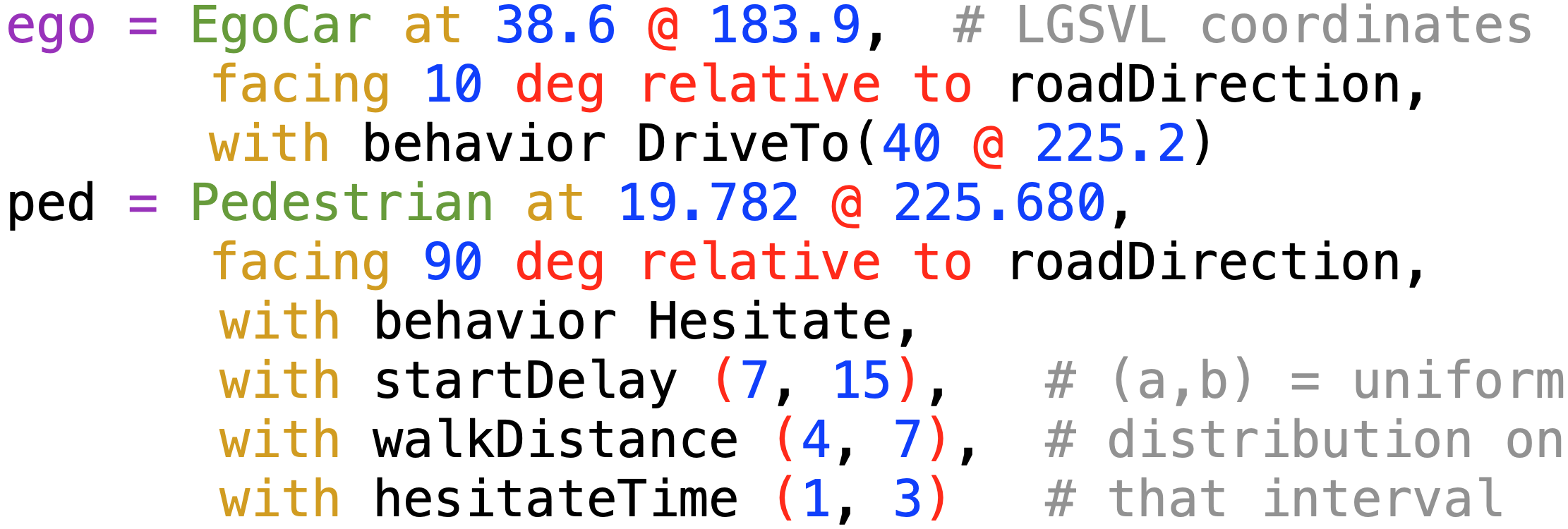}
    \caption{The (simplified) \scenic{} program $\pgm_{\Scenario}$ encoding our test scenario.}
    \label{fig:scenario-code}
\end{figure}

We encoded this scenario as the \scenic{} program shown in Fig.~\ref{fig:scenario-code}.
On lines 7--10 we specify the parameters above to have uniform distributions over appropriate ranges (e.g., $\startdelay \in (7,15)$).
The functions \texttt{DriveTo} and \texttt{Hesitate} on lines 3 and 6 specify the dynamic behavior\footnote{The \texttt{behavior} property is used by a prototype extension of \scenic{} with dynamics, which is used to define \texttt{DriveTo} and \texttt{Hesitate} and will be described in a future paper.} of the AV and the pedestrian, using API calls to Apollo and the LGSVL Simulator to command the AV to drive through the intersection and the pedestrian to walk as described above.

Finally, we defined specifications for \verifai{} to monitor during execution of our test cases.
The most important safety specification is the following: ``the AV never comes close to colliding with the pedestrian.''
We can formalize this in Metric Temporal Logic (MTL)~\cite{mtl} as $\collisionspec = \mathbf{G} (\mathit{dist} > 2.5\text{ m})$ where $\mathit{dist}$ represents the distance between the AV and the pedestrian (which we can record using the LGSVL Simulator API), and $\mathbf{G}$ is the MTL operator ``globally'', which asserts that a condition holds at every time point.
We chose a threshold of 2.5 m because we measured $\mathit{dist}$ as the distance from the center of the AV to the center of the pedestrian\footnote{In future work we plan to improve the simulator interface to measure the distance from the surface of the AV to the surface of the pedestrian.}: the distance from the center of the AV to its front bumper is 2.1 m and to its side is 0.95 m.

\subsection{Identifying Safe and Unsafe Test Cases}
\label{sec:sim:verifai}

Having defined the \scenic{} program $\pgm_\Scenario$ above, we used \verifai{} to perform falsification, sampling parameter values from the distribution $\Scenario$, running the corresponding tests in the LGSVL Simulator, and monitoring for violations of our specification $\collisionspec$. We generated 1294 test cases, of which 2\% violated $\collisionspec$.
\verifai{}'s error table stored the parameter values for all such cases, along with the quantitative satisfaction value $\rho$ of the specification; for $\collisionspec$, this is simply the minimum distance between the AV and the pedestrian over the course of the test, minus 2.5 m.
We configured \verifai{} to store the safe runs as well to distinguish robustly-safe runs from near-accident runs. The $\rho$ values help to identify marginal regions that are good candidates for testing.
Fig.~\ref{fig:startDelay-walkDistance} shows $\rho$ as a function of the start delay $\startdelay$ and the walk distance $\walkdistance$.
The darker the points, the smaller the values of $\rho$, i.e.~the closer they are to a collision.
We can see that there is no simple relation between parameter values and collisions that could be determined with a few manually-selected tests: systematic falsification was crucial for test generation.

\subsection{Test Case Selection}
\label{sec:sim:select}

\begin{figure}[tb]
    \centering
    \includegraphics[width=\columnwidth]{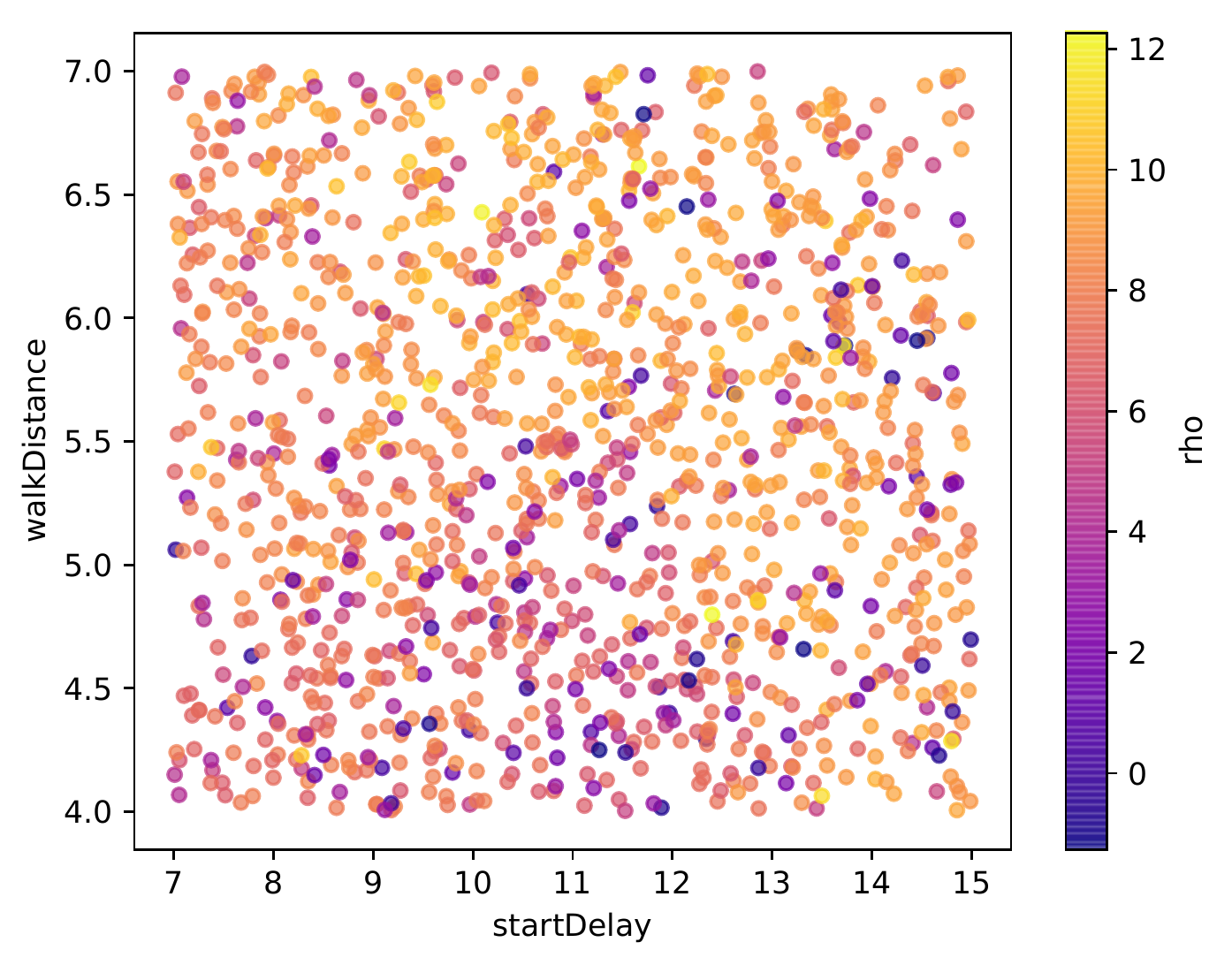}
    \caption{Satisfaction value $\rho$ of $\collisionspec$ (i.e.~the minimum distance from the AV to the pedestrian, minus 2.5 m) as a function of $\startdelay$ and $\walkdistance$. }
    \label{fig:startDelay-walkDistance}
\end{figure}

In our experiments, the parameter vector $\params$ was low-dimensional enough for direct visualization (there being only 3 parameters).
We observe in Fig.~\ref{fig:startDelay-walkDistance} that there is one main cluster of unsafe runs, in the bottom-left, and other unsafe runs towards the right for large values of $\startdelay$; however, the latter
were harder to implement due to limitations of track equipment.
Using Fig.~\ref{fig:startDelay-walkDistance} and similar plots for the hesitate time $\hesitatetime$, 
we selected values of $\params$ corresponding to three kinds of tests: failure/unsafe (F), marginally safe (M), and robustly safe/success (S). The success cases were selected from the upper-left quadrant of Fig.~\ref{fig:startDelay-walkDistance} and have a neighborhood of safe tests. The failure and marginal cases were selected from the bottom-left quadrant. The marginal cases are those that satisfy $\collisionspec$, but lie close to other failure cases; hence, implementing these cases in the real world may result in failure due to imprecision in implementing $\params$ on real hardware.
We thereby obtained 7 test cases to execute on the track as shown in Table~\ref{table:selected_test_scenarios}.

\begin{table}[t]
\caption{Track Test Cases Selected from Simulation}
\label{table:selected_test_scenarios}
\centering
{
\renewcommand{\arraystretch}{1.1}
\begin{tabular}{|c|c|c|c|c|}  
\hline
 & Hesitate & Walk & Start & Minimum \\
Test Case & Time (s) & Distance (m) & Delay (s) & Distance (m) \\
\hline \hline
F1 & 2.67 & 4.50 & 10.54 & 2.23 \\
\hline
F2 & 2.93 & 4.24 & 11.53 & 1.91 \\
\hline
M1 & 2.13 & 4.23 & 8.50 & 4.05 \\
\hline
M2 & 1.96 & 5.02 & 8.77 & 4.78 \\
\hline
M3 & 1.03 & 4.92 & 9.97 & 5.85 \\
\hline
S1 & 2.85 & 6.88 & 7.64 & 5.45 \\
\hline
S2 & 2.50 & 6.33 & 8.39 & 5.95 \\
\hline
\end{tabular}
}
\vspace*{-3mm}
\end{table}

\section{Track Testing}\label{sec:track}

\subsection{Experimental Setup}
\label{sec:track-setup}
\noindent
{\em Test AV:}
The test vehicle is a 2018 Lincoln MKZ Hybrid (shown in Fig.~\ref{fig:LG_Car}) enhanced with DataSpeed drive-by-wire functionality and several sensors including a Velodyne VLS-128 LiDAR, three Leopard Imaging AR023ZWDR USB cameras, and a Novatel PwrPak7 dual-antenna GPS/IMU with RTK correction for $\sim$2 cm position accuracy. The tests were performed using the open-source Apollo 3.5\footnote{This was the most recent Apollo version supported by our hardware.} self-driving software~\cite{apollo-www} installed on an x86 Industrial PC with an NVIDIA GTX-1080 GPU. Apollo’s perception processes data from the LiDAR sensor using GPU-accelerated deep neural networks to identify perceived obstacles.

\noindent
{\em Pedestrian Dummy and Associated Hardware:}
To implement the pedestrian at GoMentum, we used a pulley-based 4Active surfboard platform (SB)~\cite{4Active}.
The battery powered system drives a motor unit that pulls a drivable platform upon which a ``soft target'', i.e., an articulated pedestrian dummy~\cite{4Active_pedestrian}, is mounted.
The dummy is designed to have a sensor signature similar to real pedestrians.
The SB can be programmed for various types of motions, including the ``hesitating pedestrian'' trajectory used in our scenario.

\noindent
{\em Triggering Mechanisms:}
The trigger mechanism of the SB initiates the movement of the pedestrian. For repeatability of scenario testing, it is critical that the same trigger mechanism is implemented both in simulation and in real world.
We originally attempted to configure the SB to trigger automatically when a desired distance $\startdistance$ between the AV and the pedestrian is met.
However, the SB manufacturer confirmed that the SB does not support triggering based on distance threshold, and we experimentally confirmed that manual triggering based on an estimate of $\startdistance$ is not accurate.
Therefore, as described in Sec.~\ref{sec:sim:scenic}, we reparametrized our scenario in terms of a threshold \emph{delay} $\startdelay$ measured as the time elapsed from when the AV begins to move to when the pedestrian begins to move.
Although the SB hardware does not support automatic triggering based on a time delay either, we were able to implement more accurate triggering by starting a countdown timer when the AV begins to move and manually triggering the SB when the timer expired.

Setting up track tests was a tedious and time-consuming effort: it took about 8 people half a day (4 hours) to simply set up the scenario and calibrate the AV and equipment, and then another half a day to go through around 25 test runs, with each run taking 10-15 minutes.

\subsection{Test Results}

We executed the 7 test cases whose parameters are shown in Table~\ref{table:selected_test_scenarios} at GoMentum.
We performed several runs of each test scenario, obtaining 23 runs in total; these are summarized in Table~\ref{table:Quantative_Analysis}.
The highlighted rows in the table are runs which violated $\collisionspec$ (i.e., have $\rho < 0$), while the white rows satisfied $\collisionspec$.
The colors of the highlighted rows indicate the degree of unsafe behavior of the AV: red represents a collision (see Fig.~\ref{fig:LG_Car}), orange represents what we visually classified as a near-collision, and yellow violates $\collisionspec$ but is not a near-collision.
This coloring scheme brings out distinctions that are not obvious from the Table~\ref{table:Quantative_Analysis} column values.
In particular, when $\collisionspec$ is violated, the pedestrian can be approaching the car from its side or from the front: since the car is much longer than it is wide, a violation of $\collisionspec$ from a side approach is not always a (near-)collision, resulting in the yellow rows in Table~\ref{table:Quantative_Analysis}.

Both in the simulator and at GoMentum, the minimum distance is computed from the \textit{center} of the AV to the pedestrian. Hence, the minimum distance is greater than zero even though a collision occurred in F1 Run1 in Table~\ref{table:Quantative_Analysis}.
The time-to-collision (TTC) is approximated by dividing the distance between the AV and the pedestrian by the AV’s relative speed at every timestamp until either the pedestrian fully crossed the lane or the AV intersected the pedestrian path before it crossed the lane.
Videos of all our tests are available at \url{http://bit.ly/GoM_Videos}, with a visualization of Apollo's perception and planning at \url{http://bit.ly/DRV_Videos}.
We will discuss the causes of these failure cases in Sec.~\ref{subsection:why_AV_fail}.

\begin{table}[t]
\caption{Test cases selected from simulation, with corresponding runs at GoMentum. 
Runs violating $\collisionspec$ are highlighted, classified into collisions (red), near-collisions (orange), or runs which were unsafe but with a larger margin (yellow).}
\label{table:Quantative_Analysis}
\centering
{
\renewcommand{\arraystretch}{1.1}
\begin{tabular}{|c|c|c|c|}
\hline
Test Run & Minimum TTC & Minimum Distance & $\rho$\\
\hline \hline
F1 Simulation & --  & 2.23 & -0.27\\
\hline
\rowcolor{LightRed}
F1 Run1 & 2.10 & 2.06 & -0.44\\
\rowcolor{LighterYellow}
F1 Run2 & 1.27 & 2.24 & -0.26 \\
F1 Run3 & 2.97 & 4.02 & 1.52\\
F1 Run4 & 5.05 & 6.19 & 3.69 \\
\hline\hline
F2 Simulation & -- & 1.91 & -0.59\\
\hline
\rowcolor{LighterYellow}
F2 Run1 & 0.94 & 2.44 & -0.06\\
F2 Run2 & 2.70 & 3.24 & 0.74\\
\rowcolor{LighterOrange}
F2 Run3 & 1.20 & 1.58 & -0.92\\
\rowcolor{LighterOrange}
F2 Run4 & 1.05 & 2.24 & -0.26\\
\hline\hline
M1 Simulation & -- & 4.05 & 1.55\\
\hline
M1 Run1 & 6.07 & 7.20 & 4.70\\
M1 Run2 & 7.16 & 7.89 & 5.39\\
\hline
\hline
M2 Simulation & -- & 4.78 & 2.28\\
\hline
M2 Run1 & 3.24 & 3.40 & 0.90\\
M2 Run2 & 6.16 & 8.01 & 5.51\\
M2 Run3 & 9.10 & 14.38 & 11.88\\
M2 Run4 & 6.80 & 8.05 & 5.55\\
M2 Run5 & 7.69 & 8.48 & 5.98\\
\hline\hline
M3 Simulation & -- & 5.85 & 3.35\\
\hline
\rowcolor{LighterOrange}
M3 Run1 & 0.75 & 1.94 & -0.56\\
M3 Run2 & 6.00 & 6.36 & 3.86\\
M3 Run3 & 4.27 & 5.73 & 3.23\\
\hline\hline
S1 Simulation & -- & 5.45 & 2.95\\
\hline
S1 Run1 & 1.32 & 2.79 & 0.29\\
S1 Run2 & 9.72 & 8.50 & 6.00\\
S1 Run3 & 9.35 & 7.85 & 5.35\\
\hline\hline
S2 Simulation & -- & 5.95 & 3.45\\
\hline
S2 Run1 & 3.13 & 6.36 & 3.86\\
S2 Run2 & 8.66 & 9.00 & 6.50\\
\hline
\end{tabular}
}
\vspace*{-5mm}
\end{table}

\section{Data Analysis}\label{sec:analysis}

\subsection{Effectiveness of our Methodology}
\label{sec:effectiveness}

We first consider whether formal simulation was effective at designing track tests that revealed unsafe behaviors of the AV.
From Table~\ref{table:Quantative_Analysis}, we can see that this was in fact the case: out of 8 runs of the two failure tests that were identified in simulation, 5 violated $\collisionspec$ in reality, including one actual collision. 
For example, in test F1 Run1, the AV initially braked, before repeatedly inching forward while the pedestrian hesitated, ultimately colliding with it as shown in Fig.~\ref{fig:LG_Car}.
More noticeably, in 93.3\% of all (marginally) safe runs, AV satisfied $\collisionspec$ in reality. As expected, a violation case occurred in a marginally safe test, but none in safe tests. While the number of runs is small due to limited time and resources, they clearly demonstrate how simulation can be efficiently used to identify real-world failures.

On the other hand, Table~\ref{table:Quantative_Analysis} also shows that the results of a test on the track can deviate significantly from results in simulation.
For example, our first run of test case M3, which was safe in simulation, yielded a very near miss, with a minimum distance of 1.94 m (vs. 5.85 m in simulation).
The track test results also have significant variability, with test case M2 for example having minimum distances ranging widely from 3--14 m in different runs.
In the next section, we look at these discrepancies in more detail.

\subsection{The Gap Between Simulation and Reality}

There are a variety of possible sources of such discrepancies between simulation and track runs, including:
\begin{myitemize}
    \item mismatch in the initial conditions of the test (e.g.~the AV starting at a different location due to GPS error);
    \item mismatch in the dynamics of the AV or pedestrian (e.g.~incompletely-modeled physics in the simulator or imprecision in the mechanism triggering the SB);
    \item mismatch in the AV's sensory input (e.g.~reduced LiDAR point cloud density in simulation, or synthetic image rendering);
    \item timing differences due to Apollo running on different hardware in the AV and our simulation setup.
\end{myitemize}

Some of these sources of variation could be eliminated with improvements to the experimental setup which were not possible here given resource and time constraints, e.g., hardware that permits automating triggering based on $\startdelay$ or $\startdistance$ can reduce error in implementing $\params$.
However, other potential sources of error are hard to quantify: even small details of rendering, for example, could potentially change the behavior of the perception components in Apollo.
To measure the sim-to-real gap resulting from such sources, we used traces recorded from several tests to extract implemented $\params$ on the track.
We then ran a new simulation using these $\params$, which would ideally reproduce the same trace as the track test; in fact, we ran 5 identical simulations per test to assess the nondeterminism of the hardware and AV stack.

\begin{figure}[tb]
    \centering
    \includegraphics[trim={0 20px 0 0},clip,width=0.47\columnwidth]{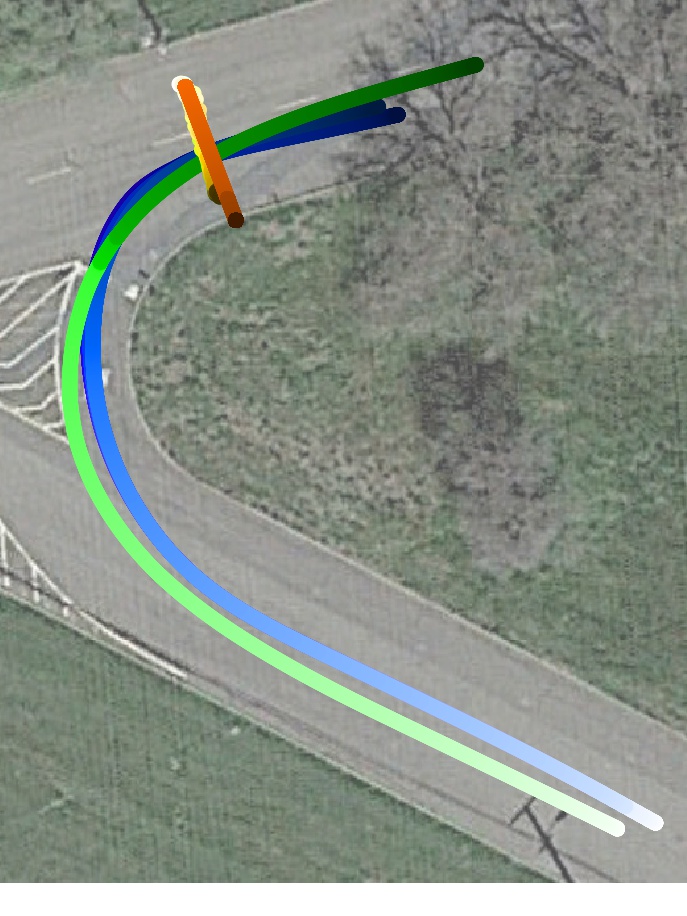}
    \hspace{0.02\columnwidth}
    \includegraphics[width=0.47\columnwidth,trim={0 20px 0 0},clip]{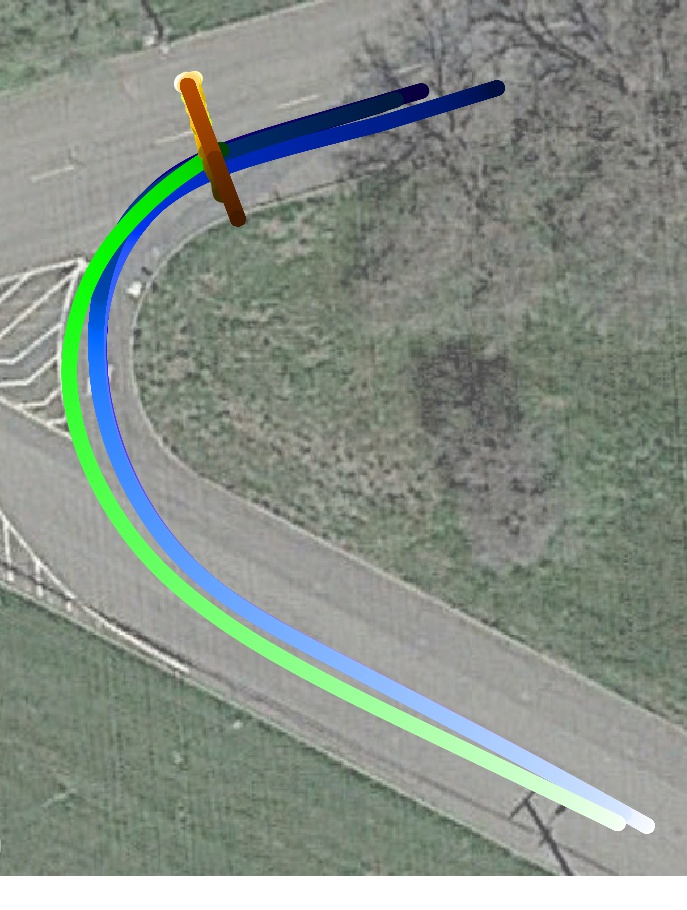}
    \caption{Trajectories from a track test and 5 resimulations respectively for the AV (green/blue) and the pedestrian (orange/yellow). Color darkness indicates time. Scenarios: S1 Run 2 (left), F1 Run 1 (right).}
    \label{fig:trajectories}
    \vspace*{-2mm}
\end{figure}

The resulting trajectories for 2 test cases are shown in Fig.~\ref{fig:trajectories}.
The simulated and real trajectories show considerable overlap but also differ: for example, although we intended the simulated AV to start from the same position as the real one, Apollo refused to drive from that position and we had to adjust it slightly.
More interestingly, the simulated AV turns more sharply than the real one, possibly due to imprecise modeling in the simulator of the effects of driving slightly uphill.
Finally, the 5 simulations are tightly-clustered but not identical, showing that even a single test case can yield a range of different simulated traces.

\begin{table}[tb]
    \caption{Avg. distances between track/resimulated AV trajectories. Measures use the $L_2$ norm with units of meters and seconds.}
    \label{tab:distance-metrics}
    \centering
    {
    \renewcommand{\arraystretch}{1.2}
    \begin{tabular}{|c|r|r|r|r|}
        \hline
        Track Test & \multicolumn{2}{c|}{Real-to-Sim} & \multicolumn{2}{c|}{Sim-to-Sim} \\
        Run & Skorokhod & DTW & Skorokhod & DTW \\
        \hline
        S1 Run2 & 5.85 & 1.09 & 1.11 & 0.17 \\
        \hline
        S3 Run3 & 5.05 & 0.91 & 2.83 & 0.24 \\
        \hline
        F1 Run1 & 10.88 & 1.39 & 3.67 & 0.37 \\
        \hline
        F1 Run3 & 5.08 & 0.90 & 3.29 & 0.26 \\
        \hline
    \end{tabular}
    }
    \vspace*{-2mm}
\end{table}

To quantify these discrepancies, taking into account not only the shape of the trajectories but also their evolution over time, we used the Skorokhod metric, which measures the worst-case deviation of two timed traces and can be used to prove conformance: a bound on how far simulated traces can be from real ones in the Skorokhod metric allows transferring temporal logic guarantees from simulation to reality~\cite{skorokhod}.
To illustrate the metric, two otherwise-identical trajectories which differ at one point by 1 m or are globally shifted by 1 s would have a Skorokhod distance of 1.
We also use the (normalized) Dynamic Time Warping (DTW) distance~\cite{dtw} to give a measure of similarity averaged over the entire trajectory rather than at the worst point.
For example, two trajectories differing by 1 m at a single point (out of many) would have a normalized DTW distance of approximately 0, vs. 1 for trajectories differing \emph{everywhere} by 1 m.

Table~\ref{tab:distance-metrics} shows these measures for the 4 track tests where we successfully logged accurate GPS trajectories.
The ``Real-to-Sim'' column shows the average distance between the real trajectory and the 5 corresponding simulations, while the ``Sim-to-Sim'' column shows the average distance among the 5 simulations.
The Skorokhod distance between the real and simulated trajectories is large, indicating deviations on the order of 5--11 meters or seconds.
This suggests that while our simulation environment was faithful enough to reality to produce qualitatively similar runs (as we saw in Sec.~\ref{sec:effectiveness}), it is not yet accurate enough to enable deriving guarantees on the real system purely through simulation.

Although the simulations are much closer to each other than to the original track test, they still show substantial variation (e.g.,~the resimulations of F1 Run 1 have large Skorokhod distance).
As the LGSVL Simulator is deterministic, this shows that either the asynchronous interface to Apollo or nondeterminism within Apollo itself can produce significantly different behavior on identical test cases.
Our methodology could likely be improved by taking this nondeterminism into account: tests with lower variance are more likely to be reproducible on the track, while tests with high variance may indicate undesirable sensitivity of the AV stack.

\subsection{Why Did the Autonomous Vehicle Fail?}\label{subsection:why_AV_fail}

\begin{table}[t]
\caption{Hypothesized causes of the observed unsafe behavior.}
\label{table:Qualitative_Analysis}
\centering
{
\begin{tabular}{|c|c|c|c|}  
\hline
Test Case & Perception Fail. & Prediction Fail.
& Planning Fail.\\ \hline\hline
F1 Run1 & \checkmark & -- & -- \\
F1 Run2 & -- & -- & \checkmark \\
\hline\hline
F2 Run1 & -- & \checkmark & -- \\
F2 Run3 & -- & -- & \checkmark \\
F2 Run4 & -- & -- & \checkmark \\
\hline\hline
M3 Run1 & -- & -- & \checkmark \\
\hline
\end{tabular}
}
\end{table}

Finally, by replaying our track data in Dreamview, a tool to visualize Apollo's perception, prediction, and planning, we identified several types of failures that led to unsafe behavior, summarized in Table~\ref{table:Qualitative_Analysis}.
The simplest type was a perception failure, where Apollo failed to detect the pedestrian for at least 1 s: this was responsible for the crash in Fig.~\ref{fig:LG_Car}.
The most common failure was unsafe planning, where Apollo alternated between yielding and overtaking the pedestrian.
Most interestingly, we also observed a case of prediction failure, where the AV incorrectly predicted that the pedestrian would walk around the AV and that moving forward was, therefore, safe.
While our goal is not to find fault with Apollo (recall also that we could only run version 3.5 from January 2019 on our hardware), our results
illustrate how our methodology can help to find and debug failure cases.

\section{Conclusion}\label{sec:concl}
We presented a formal methods approach to scenario-based test generation for autonomous vehicles in simulation, and to the selection and execution of road tests, leveraging the \scenic language and the \verifai toolkit.
We demonstrated that a formal simulation approach can be effective at identifying relevant tests for track testing with a real AV.
We also compared time-series data recorded in simulation and on the track, both quantitatively and qualitatively.
Our data and analysis scripts can be found online~\cite{data-repo}.

There are several directions for future work, including evaluating our methodology on more complex higher-dimensional scenarios, performing more detailed automated analysis of failures in perception, planning, or prediction, bridging the sim-to-real gap with further improvements to simulation technology, and developing more sophisticated track test equipment that can better match simulation.




%

\section*{Acknowledgments}

The authors thank Hadi Ravanbakhsh for substantial assistance with early experiments, Francis Indaheng for assistance with the \scenic-LGSVL interface, Jyotirmoy Deshmukh for providing code to compute the Skorokhod metric, and the anonymous reviewers for their helpful comments.


\bibliographystyle{IEEEtran}
\bibliography{refs}

\end{document}